# Free-space Excitation of Resonant Cavities formed from Cloaking Metamaterial

Edward P. Furlani[*] and Alexander Baev

*The Institute for Lasers, Photonics and Biophotonics*

*State University of New York at Buffalo*

*Buffalo, New York 14260, USA*

[*]*Corresponding author: efurlani@buffalo.edu*

We propose a new class of resonant electromagnetic structures, and study their response to free-space illumination. The structures consist of partial cylindrical shells that have cloaking material properties proposed by Pendry *et al*. These metamaterial shells have apertures that allow the propagation of incident irradiation into an interior resonant cavity. We use full wave time-harmonic analysis to study the field distribution inside the cavity, and show that an analogue of Whispering Gallery Modes (WGMs) can be efficiently excited via free-space illumination.

Recently, Pendry *et al*. have proposed an interesting method for designing media that enable electromagnetic "cloaking" of spatial regions [1]. The method involves the use of a coordinate transformation that mathematically "squeezes" the cloaked region into a surrounding shell. The cloaking material properties of the shell are determined from the transformation itself, which is mapped to anisotropic permittivity and permeability tensors. The shell, with these properties, shields both its interior cloaked volume from external irradiation, and its exterior domain from fields generated from within its interior. The concept of cloaking a region using custom tailored media has attracted substantial interest especially in fields such as stealth technology. Moreover, the development of such media should benefit from the significant progress made in the field of metamaterials [2-12]. In this regard, a crude cloaking system has already been demonstrated at microwave frequencies using artificially structured materials [13]. To date, almost all of the research on cloaking media has focused on an analysis of fully cloaked cylindrical and spherical regions [14-19]. In this paper, we present a new application of cloaking metamaterial that involves the formation of local resonant cavities.

We propose a new class of resonant structures, and study their response to free-space illumination. These structures consist of partial cylindrical shells of cloaking metamaterial with apertures that allow the penetration of incident irradiation into an interior resonant cavity (Fig. 1). Throughout this paper we assume that the cloaking structures are in free-space. We consider microscale cavities with resonant frequencies in the optical range. There is considerable interest in such systems for applications such as enhanced Raman spectroscopy (CERS) [20,21], lasing [22,23], and ultra-sensitive bio-detection [24-27]. Resonant micro-cavities confine the electromagnetic field, which results in local field enhancement. The quality factor $Q$ of such structures can be quite large [28], which enables the detection of small perturbations via a measurable spectral shift in the resonant modes commonly referred to as Whispering Gallery Modes (WGMs) or Morphology Dependent Resonances (MDRs) [24-27]. High $Q$ resonant cavities are especially of interest for biosensing as they can detect small local changes in the index of refraction, which can be caused by the presence of a target biomaterial. One problem with conventional dielectric WGM-based sensors is that free-space excitation of the resonant modes tends to be highly inefficient [29]. While tapered waveguides or prisms enable efficient evanescent field coupling to these modes, they are not compatible with free-space excitation [30, 31].

In the resonant structures that we propose the resonant modes can be excited using free-space illumination. We demonstrate the resonant behavior of the structures using full-wave time-harmonic field analysis. We begin the analysis with a briefly review of cylindrical cloaking. Consider a cylindrical region of radius $R_2$ that is centered at the origin in the *x-y* plane. Following Pendry *et al.*, we apply the following coordinate transformation, which "squeezes" the entire cylindrical region $0 \leq r < R_2$ into the annulus $R_1 < r < R_2$,

$$\begin{aligned} r' &= a + \alpha r, \\ \theta' &= \theta, \\ z' &= z, \end{aligned} \quad (1)$$

where $\alpha = (R_2 - R_1)/R_2$ [1]. This transformation is mapped to "cloaking" material tensors $\bar{\bar{\varepsilon}}_c$ and $\bar{\bar{\mu}}_c$ as follows [16,19],

$$\bar{\bar{\varepsilon}}_c = \bar{\bar{\mu}}_c = \begin{bmatrix} \dfrac{r'^2 + R_1(R_1 - 2r')\cos^2(\theta')}{r'(r' - R_1)} & \dfrac{R_1(R_1 - 2r')\sin(\theta')\cos(\theta')}{r'(r' - R_1)} & 0 \\ \dfrac{R_1(R_1 - 2r')\sin(\theta')\cos(\theta')}{r'(r' - R_1)} & \dfrac{r'^2 + R_1(R_1 - 2r')\sin^2(\theta')}{r'(r' - R_1)} & 0 \\ 0 & 0 & \dfrac{r' - R_1}{\alpha^2 r'} \end{bmatrix}. \quad (2)$$

Eq. (2) gives the Cartesian components of the material tensors in terms of the radial coordinates $(r', \theta')$, e.g., $\varepsilon_{c,xx} = \left(r'^2 + R_1(R_1 - 2r')\cos^2(\theta')\right)\big/\left(r'(r' - R_1)\right)$. These components are functions of the transformed Cartesian coordinates $(x', y')$ through the relations $r' = \sqrt{(x')^2 + (y')^2}$ and $\theta' = \tan^{-1}(y'/x')$, i.e. $\bar{\bar{\varepsilon}}_c(x', y') = \bar{\bar{\mu}}_c(x', y')$. To analyze a cloaked cylinder ($r < R_1$), we solve Maxwell's equations in the $x'$-$y'$ plane using the material properties $\bar{\bar{\varepsilon}}_c$ and $\bar{\bar{\mu}}_c$ for the cylindrical shell cloaking material, and $\bar{\bar{\varepsilon}} = \bar{\bar{\mu}} = I$ for all other regions, where $I$ is the 3x3 identity matrix.

We study the resonant behavior of the partial cloaking shell structure shown in Fig. 1. We assume that the shell has inner and outer radii of $R_1$ = 3 μm and $R_2$ = 4 μm, respectively, and an aperture that allows incident light to penetrate into the central resonant cavity, which is a region of free-space. We perform a 2D full-wave simulation of this system using the COMSOL Multiphysics FEA-based RF solver. The computational domain is 14.4 μm high and 18.4 μm wide, which includes surrounding perfectly matched layers (PMLs) (Fig. 1). We illuminate the structure with a transverse-electric (TE) polarized time-harmonic Gaussian wave, which is generated using a surface current boundary condition of the form $J_z = 1 \times 10^4 \exp(-y \times 10^6)^2$ V/m.

It is instructive to first examine the field distribution for both free-space illumination (i.e. no cloaking material), and illumination of a complete cloaking shell.

The wavelength of the incident TE field is set to $\lambda = 1004$ nm. In the free-space analysis, which is shown in Fig. 2a, a cylindrical shell region is present, but its material properties are set to unity ($\bar{\bar{\varepsilon}}_c = \bar{\bar{\mu}}_c = I$). Note that the field distribution expands with distance from the source as expected due to the Gaussian spatial distribution of the imposed surface current. However, when the cloaking shell is present the field is distorted within the shell so as to shield and cloak its interior as shown in Fig. 2b, i.e. there is no field inside the shell and the field distribution beyond the shell approaches the free-space case.

We now consider the partial shell structure of Fig. 1. We study the response of this structure under TE illumination at two different wavelengths; first at $\lambda = 1004$ nm, and then at $\lambda = 780$ nm. The FEA for all of the partial shell analysis below was performed using 100,288 Lagrange cubic elements. The field distribution for $\lambda = 1004$ nm is shown in Fig. 3a. Note that at this wavelength an enhanced WGM field pattern forms along the outer circumference of the inner cavity. This is due to the difference in the index of refraction between the partially cloaked interior and the surrounding shell at the interface $R_1$, which gives rise to total internal reflection. In this case, the resonant mode is of the first order with the mode number of 14. The wavelength, $\lambda$, of an $m$-th resonant mode inside a dielectric cavity with the radius $R_1$ can be estimated using the following relation:

$$2\pi R_1 = m\lambda_m. \tag{3}$$

According to Eq. (3), the effective radius of the resonant cavity is 2.237 μm at $\lambda = 1004$ nm, which differs from the physical radius of the cavity, i.e. 3 μm. Apparently, the "effective" resonator has the radius defined by location of the maximum field value of the mode. This is consistent with the field pattern obtained using the $\lambda = 780$ nm illumination, which excites a second order mode with azimuthal number 15 as shown in Fig. 3b. In this case the effective resonator has the radius of 1.86 μm according to Eq. (3), which is also the radius of the inner edge of the mode where the field takes its maximum value. In our study, we performed additional simulations for this geometry using a variety of wavelengths and found that the resonant cavity will support a multitude of modes of various configurations. We also performed a series of simulations in which we varied the angular extent (i.e., size) of the wedge-shaped aperture, and found that this affected both the wavelength at which resonance occurred and the mode structure, as expected.

It should be noted that the proposed cloaking metamaterial structures are not unique in their ability to sustain resonant modes via free-space illumination. Indeed, it is well known that metallic conductors can be used to from such resonant electromagnetic cavities as well. However, the mode structures for these materials are quite different than those obtained using cloaking metamaterials. This is demonstrated in Fig. 4, which shows the modes that obtain in a partial shell of a perfect conductor with an inner radius of $R_1 = 3$ μm as above. The shell is infinitely thin, and we model it as a perfect electric conductor (PEC) boundary condition in free-space. The source parameters, computational domain and mesh are as defined above. The outline of the partial metamaterial shell appears in the computational domain, but its bulk electrical properties are set to free-space values. We illuminate the PEC shell with wavelengths of $\lambda = 1004$ and 780 nm, and the mode structures for these wavelengths are shown in Figs. 4a and c, respectively. Note that PEC cavity field patterns are completely different than those that obtain inside the cloaking metamaterial structure. This is because the electrical properties, and hence EM field conditions, at the boundary of the resonant cavity are different for the two materials. Specifically, $\varepsilon_r = \mu_r = 0$ and $\varepsilon_\theta = \mu_\theta = \infty$ at the boundary of the metamaterial cavity, whereas the conductivity is infinite ($\sigma = \infty$) for a PEC, and hence only a normal component of the E field can be sustained on this boundary.

Lastly, we study the performance of lossy partial shell structures [16-18]. Specifically, we introduce a loss term into each component of the permittivity and permeability tensors, and then compute the field distribution to see if the resonant modes identified above are still preserved. We study the response of the structure at both resonant wavelengths λ = 1004 nm and 780 nm, and for two different values of loss tangent [16,18]; tan(δ) = 0.1 and 0.5. The response at λ = 1004 nm is shown in Figs. 5a,b. Note that the resonant mode is still present, and that the forward propagating field is attenuated due to absorption. A similar response is observed at λ = 780 nm, which is shown in Figs. 5c,d. Thus, lossy structures would still be suitable for resonant mode applications. We also note that we studied the response of the partial shell structure under simple plane-wave illumination using a scattering field formulism. We computed the field at λ = 1004 nm and 780 nm with and with loss. We observed essentially the same behavior as above in all cases.

In summary, we have proposed a new class of resonant structures that are formed using cloaking metamaterial. These structures can be excited using free-space illumination. They are capable of sustaining a multitude of distinct and unique resonant mode configurations. The resonant modes can be tuned by adjusting the illumination wavelength, and they are sensitive to the size of the aperture and the refractive index of the material in a cavity. Thus, if cloaking materials can be realized in practice, the proposed resonant structures could be useful for applications such as nanoscale bio-sensing, where the presence of a biomaterial changes the local index of refraction.

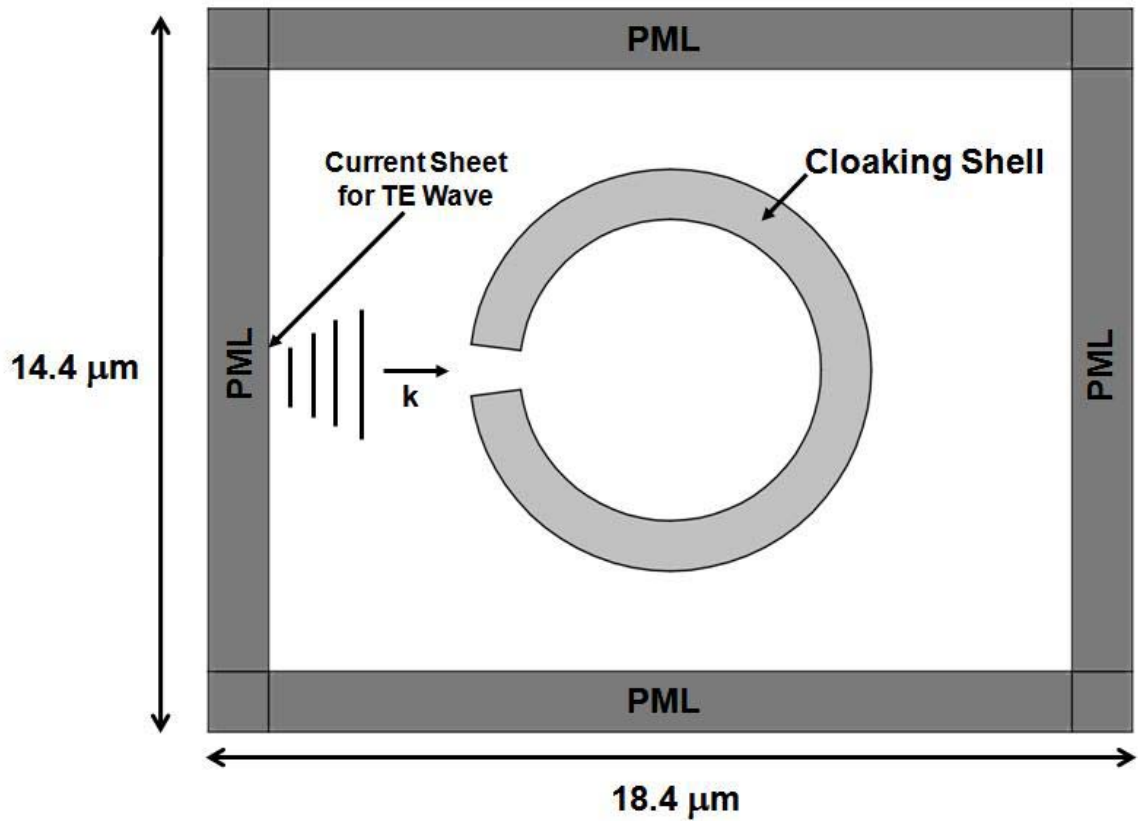

Fig. 1 Partial cylindrical shell of cloaking metamaterial illuminated by a Gaussian TE wave.

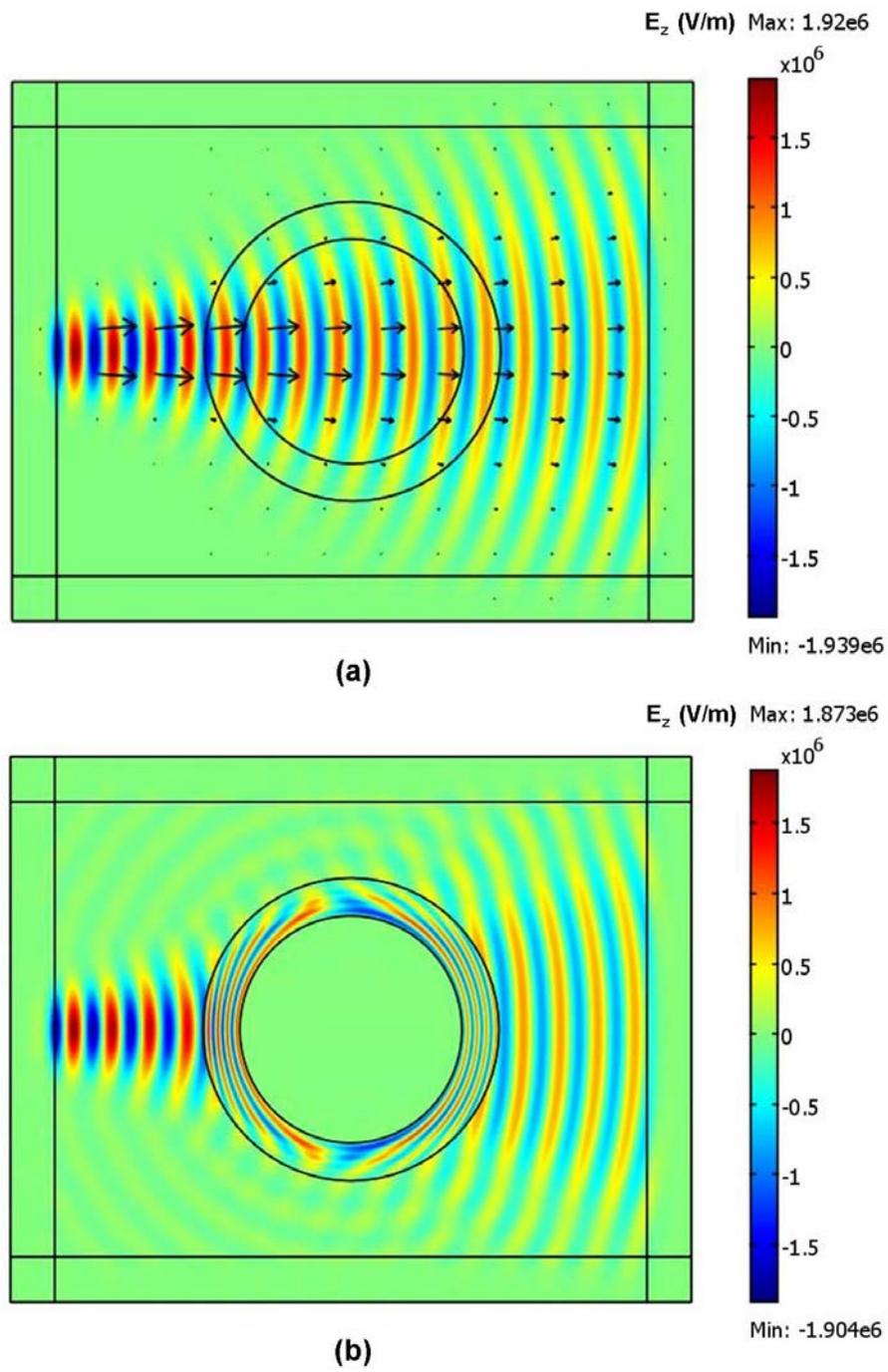

Fig. 2: Full-wave time-harmonic field analysis ($\lambda$ = 1004 nm): (a) $E_z$ and time-averaged power flow vectors for free-space propagation, and (b) $E_z$ for fully cloaked cylinder.

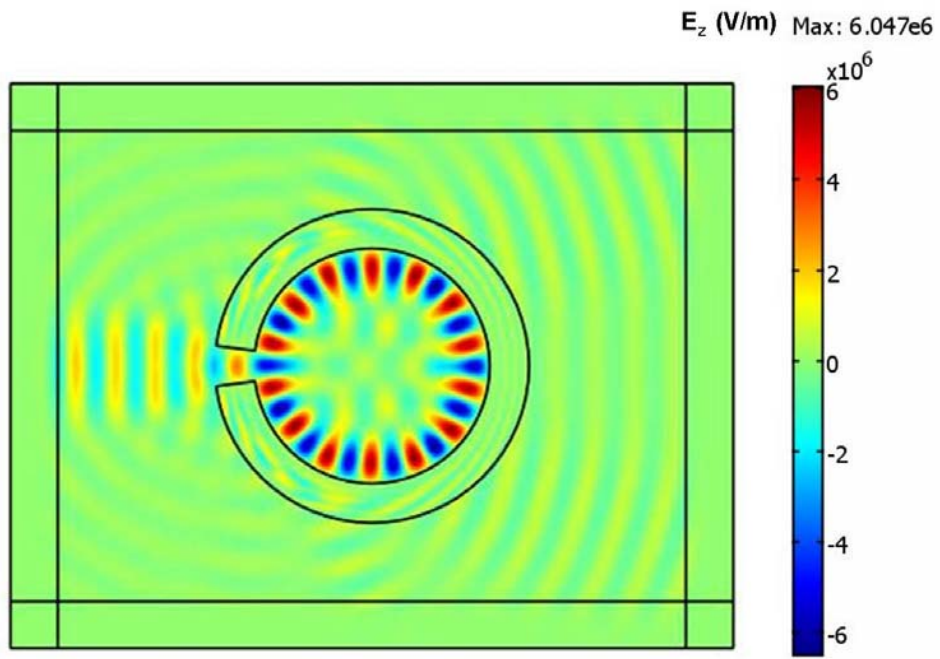

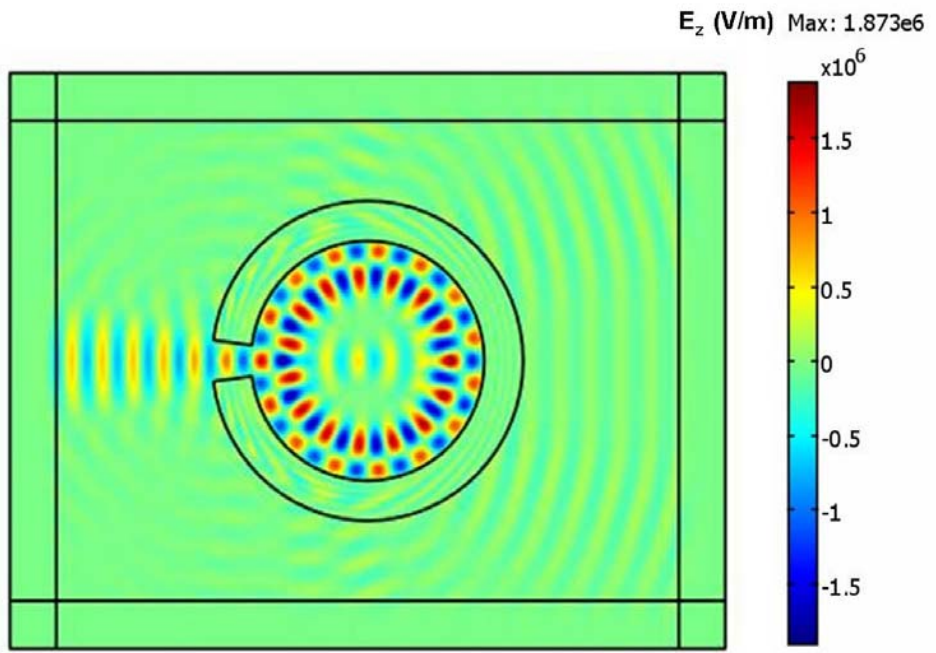

Fig. 3: Full-wave analysis ($E_z$) of partial cloaking shell at different wavelengths: (a) $\lambda =$ 1004 nm, and (b) $\lambda =$ 780 nm.

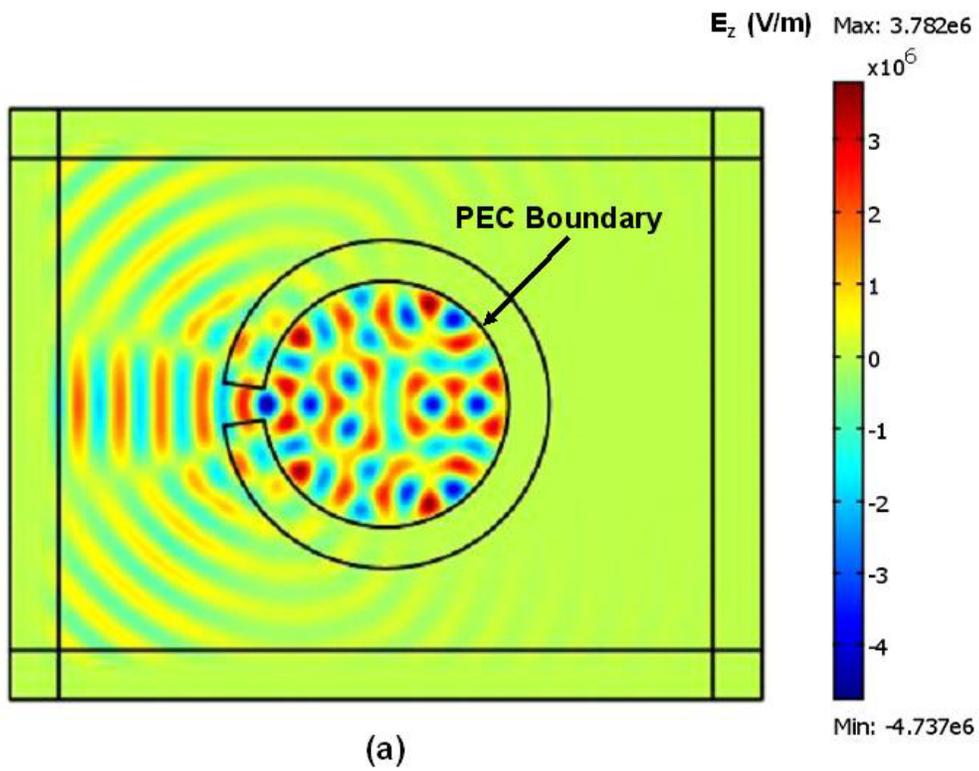

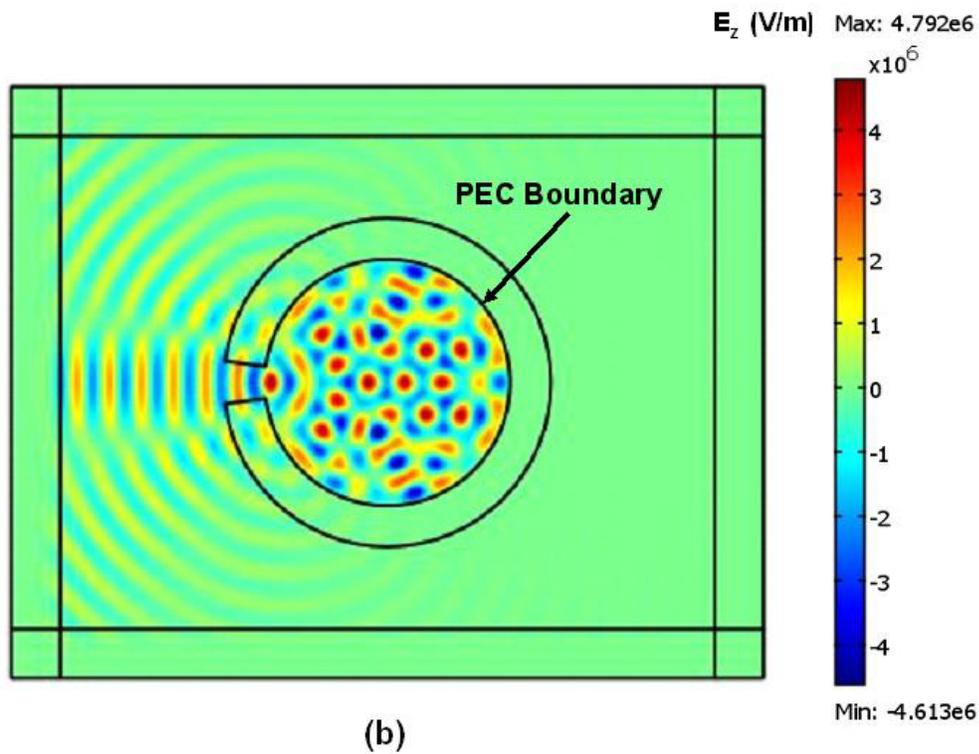

Fig. 4: Full-wave analysis ($E_z$) of partial PEC shell forming a resonant cavity in free space and illuminated at different wavelengths: (a) λ = 1004 nm, and (b) λ = 780 nm.

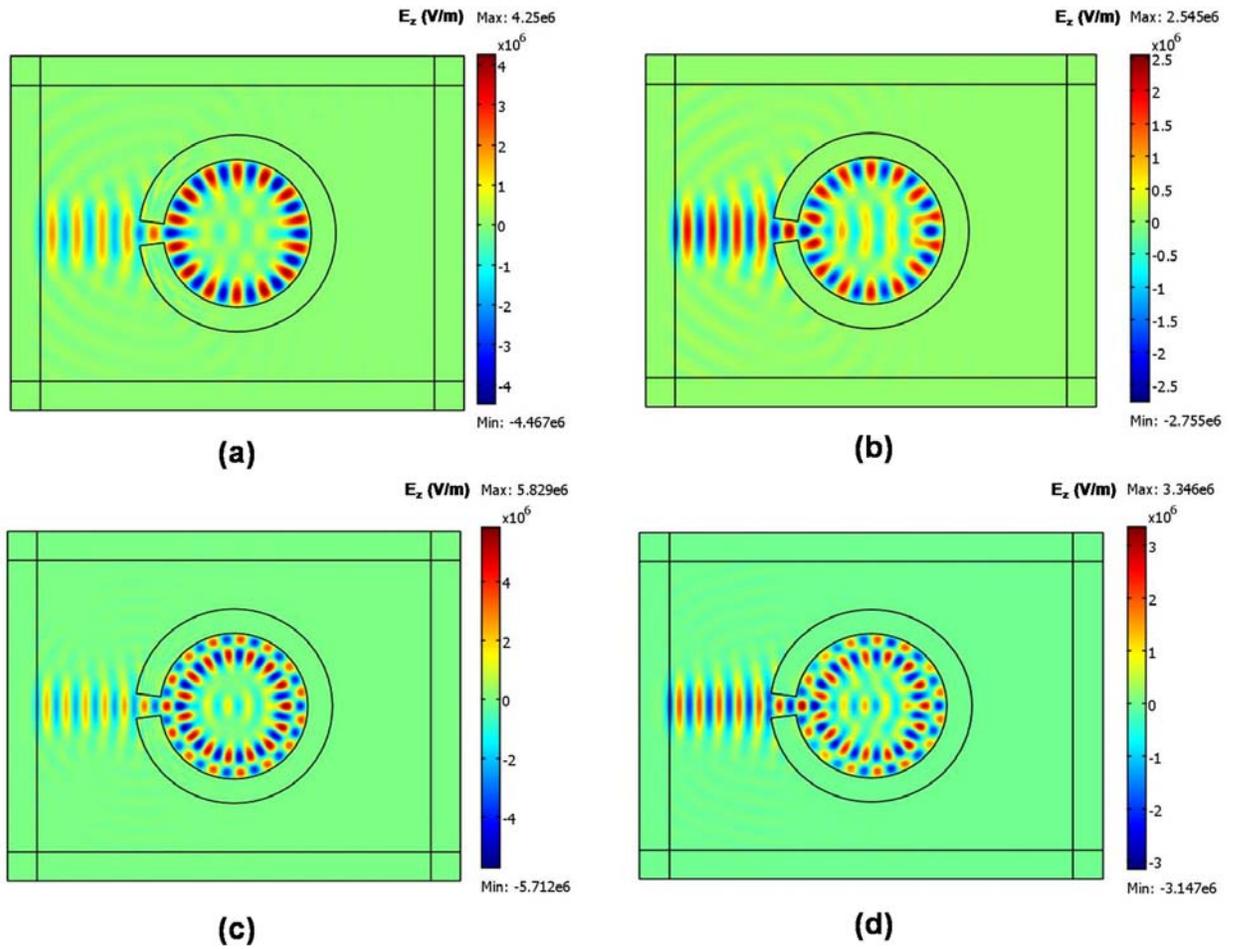

Fig. 5: Field distribution ($E_z$) of a lossy partial shell of cloaking metamaterial (loss tangent introduced into each component of the permittivity and permeability tensors): (a) $\lambda = 1004$ nm, $\tan(\delta) = 0.1$, (b) $\lambda = 1004$ nm, $\tan(\delta) = 0.5$, (c) $\lambda = 780$ nm, $\tan(\delta) = 0.1$, and (d) $\lambda = 780$ nm $\tan(\delta) = 0.5$.